\documentclass[10pt, conference]{IEEEtran}
\usepackage[utf8]{inputenc}
\usepackage{amssymb,amstext,amsmath,amsfonts}
\usepackage{algorithmic}
\usepackage{epsfig}
\usepackage{graphicx}
\usepackage{subfigure}
\usepackage{url}

\usepackage{multicol,multirow}
\usepackage{bbm}
\usepackage{nicefrac}
\usepackage{xspace}
\usepackage{acronym}

\usepackage{color}

\usepackage{booktabs}

\usepackage{longtable}

\DeclareMathOperator*{\argmax}{\arg\!\max}
\usepackage{capt-of}
\let\llncssubparagraph\subparagraph
\let\subparagraph\paragraph
\usepackage{titlesec}
\let\subparagraph\llncssubparagraph
\titlespacing{\section}{0pt}{1.5ex}{0.5ex}
\titlespacing{\subsection}{0pt}{1.5ex}{0.5ex}
\begin{document}
\title{Towards Real-time Customer Experience Prediction for Telecommunication Operators}
\author{
\IEEEauthorblockN{Ernesto Diaz-Aviles, Fabio Pinelli, Karol Lynch, Zubair Nabi,}
\IEEEauthorblockN{Yiannis Gkoufas, Eric Bouillet, and Francesco Calabrese}
\IEEEauthorblockA{
IBM Research -- Ireland\\
\texttt{\{e.diaz-aviles, fabiopin, karol\_lynch, zubairn,}
}
\IEEEauthorblockA{
\texttt{yiannisg, bouillet, fcalabre\}@ie.ibm.com}
}\\
\IEEEauthorblockN{Eoin Coughlan and Peter Holland}
\IEEEauthorblockA{IBM Now Factory -- Ireland\\
\texttt{\{eoin.coughlan, peter.holland\}@ie.ibm.com}
}\\
\IEEEauthorblockN{Jason Salzwedel}
\IEEEauthorblockA{Vodacom -- South Africa\\
\texttt{jason.salzwedel@vodacom.co.za}
}\\
} 

\maketitle              

\begin{abstract}
Telecommunications operators (telcos) traditional sources of income, voice and SMS, are shrinking due to customers using over-the-top (OTT) applications such as WhatsApp or Viber.  In this challenging environment it is critical for telcos to maintain or grow their market share, by providing users with as good an experience as possible on their network.
But the task of extracting customer insights from the vast amounts of data collected by telcos is growing in complexity and scale everey day. How can we measure and predict the quality of a user's experience on a telco network in real-time? That is the problem that we address in this paper.
We present an approach to capture, in (near) real-time, the mobile customer experience in order to assess which conditions lead the user to place a call to a telco's customer care center. To this end, we follow a supervised learning approach for prediction and train our \emph{Restricted Random Forest} model using, as a proxy for bad experience, the observed customer transactions in the telco data feed before the user places a call to a customer care center. 
We evaluate our approach using a rich dataset provided by a major African telecommunication's company and a novel big data architecture for both the training and scoring of predictive models. Our empirical study shows our solution to be effective at predicting user experience by inferring if a customer will place a call based on his current context.

These promising results open new possibilities for improved customer service, which will help telcos to reduce churn rates and improve customer experience, both factors that directly impact their revenue growth.
%
%
%
\end{abstract}


\section{Introduction}
\label{sec:introduction}
The number of mobile-cellular subscriptions worldwide is approaching the number of people on earth, with the developing countries accounting for over three quarters of the world's total~\cite{itu2014}. We are becoming more and more dependent on our mobile phones -- GPS navigation, voice and text over data, and social media exchanges are just a few examples. We demand to be always online since our work and leisure activities are impacted otherwise. Telcos struggle to meet these high demands in a market where traditional voice or text plans are disappearing in favor of data services supporting a diverse range of mobile apps. 

%

A clear up-to-date understanding of customer experience and satisfaction is a key competitive advantage for telcos~\cite{2013:forbes}. However, telcos face the challenge of dealing with large amounts of information generated by the mobile users every second. For example, mobile data traffic is forecasted to reach 24.3 Exabytes per month by 2019, which corresponds to a \mbox{10-fold} growth from 2014 to 2019~\cite{cisco_vni_stats}. Such large and fast mobile data makes it harder for telcos to extract customer insights in a timely enough manner to react to potential causes of poor customer experience.

One option the operators have is to directly ask their subscribers using surveys but this mechanism, although reliable and widely used in the industry, is infeasible for real-time or large scale settings. Ideally, if telcos were able to measure customer satisfaction at any point in time and identify potential causes of poor customer experience, it would be easier for them to address such issues promptly, before they deteriorate and impact a larger number of subscribers. 

\begin{figure}[!ht]
\centering
\includegraphics[width=\columnwidth]{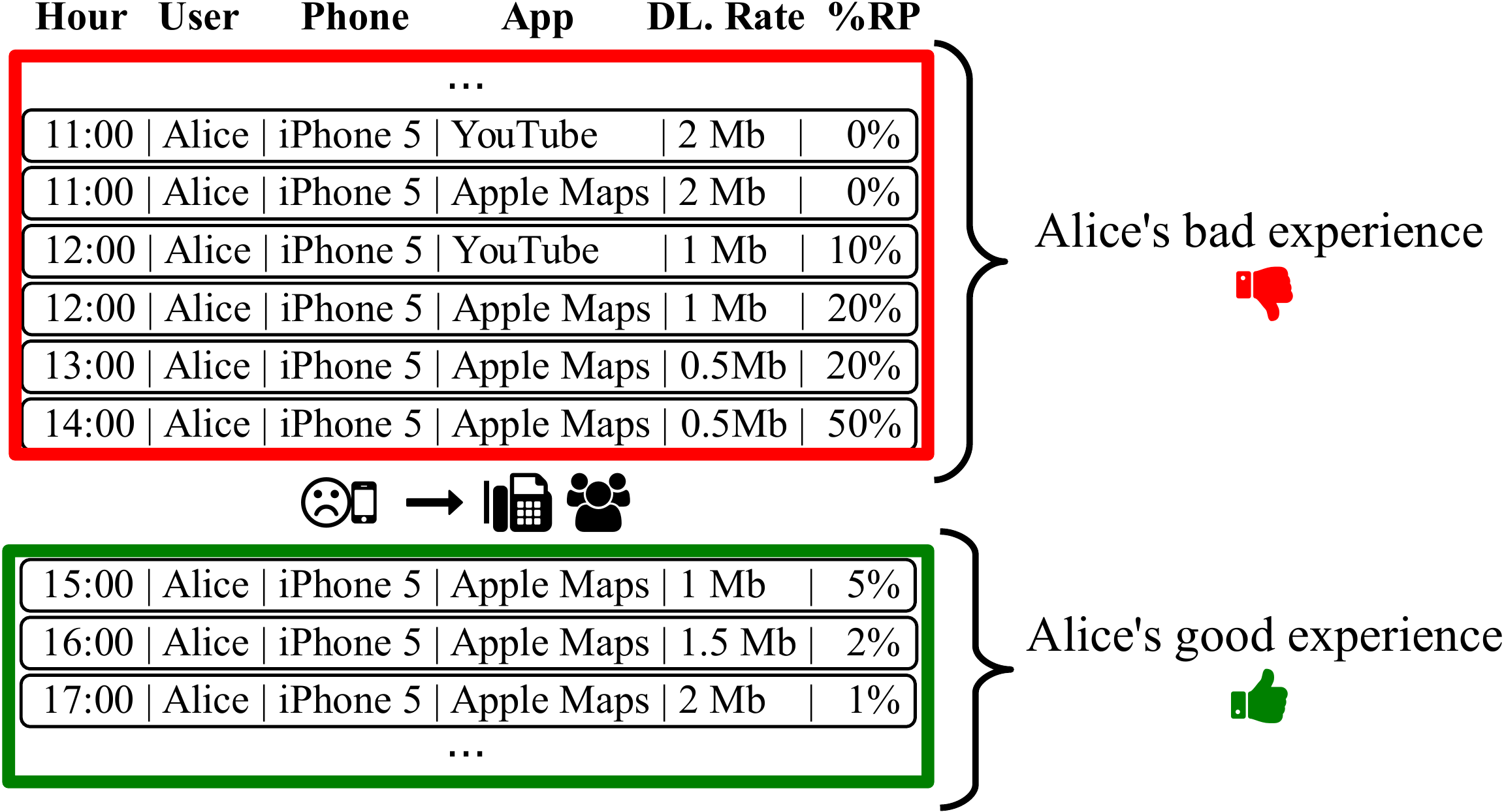}
\caption{User Alice's Data Feed transactions are illustrated here by a vector that includes the hour of the day, the user id, her device model, the app she is interacting with, the download rate, and the percentage of retransmitted packets (\%RP). One approach, explored in our Experiment~I, is to consider that all Alice's transactions before her call at 14:30 to the care center represent her bad experience. The telco's support agents solved Alice's problem and all transactions thereafter are considered a good experience.}
\label{fig:exp1_labeling}
\end{figure}

In this paper, we propose an alternative approach to infer the experience of every telco customer in (near) real-time by mining data already available to the telcom operator. This information consists of anonymized mobile phone Internet usage data associated to device location and specific app status (e.g., download rate, retransmitted packets, throughput) as illustrated in Figure~\ref{fig:exp1_labeling}. 

We make the assumption that a user's bad experience can be inferred by looking at the network performance for her use of specific mobile applications and services. However, it is very difficult to determine if such network performance is indeed perceived as an overall \emph{good} or \emph{bad} experience by the user. To tackle this challenge, we test the hypothesis that poor customer experience results in calls to the telco's care center. For instance, if a user is interacting with a maps app and this is very slow, she might decide to call the telco's care center to complain about low mobile data speed (cf. Figure~\ref{fig:exp1_labeling}).

While calls placed to customer support centers are not the only way for customers to explicitly report about bad experience, e.g., other ways include providing a low rating in Net Promoter Score (NPS) surveys or churning, we have chosen to use this valuable source of user feedback since it is much more abundant (usually an order of magnitude higher) than survey-based input. Moreover, if we are able to predict a customer care call based on network performance, we can help telcos identifying the reasons behind the complaints (e.g., a problem with a cell tower) and proactively suggest ways to solve the issue before this is experienced by a large number of users. In total we define our problem as follows.

\emph{Customer Experience Prediction Problem}. Given a set of mobile phone usage transactions, the problem of customer experience prediction is to find out if the current conditions, i.e., context, will lead the user to place a call to the customer care center in the near future.

We cast this problem as a binary prediction task, i.e., will the user call the care center or not given his current mobile experience, in which we train an ensemble of decision trees using different dimensions of user context. Decision trees help us to ease interpretability of the reasons causing bad user experience (which is an important aspect for real applications dealing with human interaction) and a model ensemble, which we refer to as a Restricted Random Forest (RRF), allows us to obtain better predictive performance.

We implemented our solution as a proof-of-concept for a major telecommunications company in Africa and deployed it in the telco's infrastructure to evaluate it using a rich dataset provided by the company. A key finding of this initial study, on the historical data provided, is that our approach was able to predict a problem with Apple Maps (in particular due to a high percentage of retransmitted packets) that resulted in many customer care calls, both during the monitored days, and for two subsequent weeks, as confirmed by the telco. If the telco had deployed the proposed solution in a real-time manner, it would have been able to identify such an issue faster, and proactively communicate it to customers thus reducing the number of customer care calls placed due to this problem. In total, the contributions of this work are as follows:

--~We propose a novel approach for telecommunication operators to measure customer satisfaction in (near) real time, which is capable of identifying potential causes of a poor customer experience by leveraging readily available data.

--~We report interesting insights on what users' context dimensions impact their mobile experience.

--~We empirically show the potential of our approach to assess the user mobile experience for the telco's entire customer base without resorting to expensive and time consuming survey-like strategies.
%
%
%
%
%
%
%

\section{Related Work}
\label{sec:related_work}

Machine learning and data mining have been widely applied to customer relationship management \cite{ngai2009application}.
There has been a large amount of work within this community around churn prediction \cite{hung2006applying} and different strategies have been proposed, e.g., based on demographics and call-behavior \cite{wei2002turning}. 

More recently, social network information has been shown to have a positive impact on churn prediction \cite{richter2010predicting,rowe2013mining}. Different techniques have been used for this task. In \cite{mozer2000predicting} logistic regression, decision trees, neural networks, and boosting were used and in \cite{zhao2005customer} support vector machines have been proposed.

Mobile data usage has been shown to be useful to correlate also with other aspects of human life, such as urban dynamics \cite{4287441} or crime \cite{Bogomolov:2014}. Previous work has shown the practical usefulness of decision trees to predict crime hotspots from mobile data \cite{Bogomolov:2014}. On a different angle, approaches have been tried to mine service center call records to better characterize service requests \cite{tan2000textual}.

However, to the best of our knowledge, no work so far has tried to address the (near) real-time customer experience prediction problem (in particular of call center calls) by making use of customer mobile Internet usage data.
\begin{table}[!tb]
\caption{2G and 3G Dataset Statistics.}
\label{tab:data_stats}
\centering
\scalebox{0.85}{
\begin{tabular}{l c r}
\toprule
Time span & : & 2014-08-08 to 2014-08-12 \\
Number of data feed transactions & : & 816,362,972 ($\sim$ 816.4M) \\
Total number of unique mobile users & : & 1,901,612 ($\sim$ 1.9M) \\
Number of unique users that call the care center & : & 63,594 ($\sim$ 63.6K) \\
Number of calls to care center & : & 107,459 ($\sim$ 107.5K) \\
\bottomrule
\end{tabular}
}
\end{table}
\section{Mobile Phone Usage and\\Customer Care Dataset}
\label{sec:data}
In this section, we describe the anonymized data sample used in our study and share insights from its exploration. The dataset consists of two main sources of information namely Data Feeds and a log of Customer Care Calls, which are detailed as follows. Table~\ref{tab:data_stats} summarizes the dataset statistics.
\begin{figure*}[!t]
\centering
\subfigure[\scriptsize Download data (GB).]{
	\frame{\includegraphics[width=0.31\linewidth]{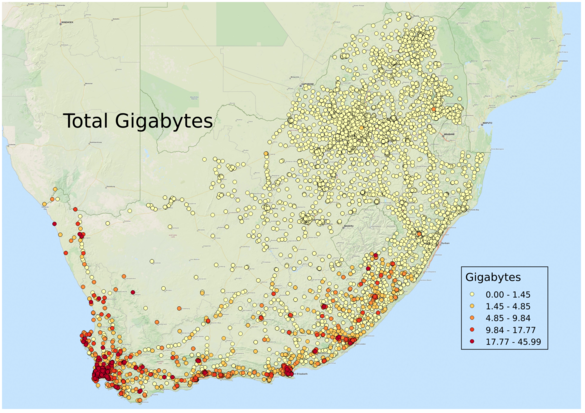}}
	\label{fig:volume_map}
	}~
	\subfigure[\scriptsize Retransmitted packets (\%).]{
	\frame{\includegraphics[width=0.31\linewidth]{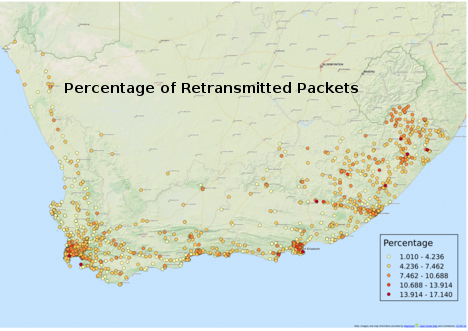}}
	\label{fig:map_retransmitted}
	}~
	\subfigure[\scriptsize Callers to care center (\%).]{
	\frame{\includegraphics[width=0.31\linewidth]{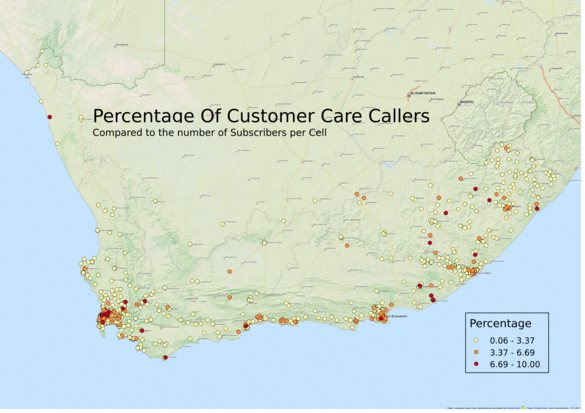}}
	\label{fig:ccc_map}
	}
\caption{Spatial distribution.}
\label{fig:dataset_exploration_geo}
\end{figure*}

\textbf{--~Data Feeds}.~This dataset is collected in real-time using a probe installed within the telecom operator's infrastructure. The data is aggregated on-the-fly to generate \emph{Data Feeds}, which encapsulate low-level summary information using customer-centric Internet measurements for different aggregation time periods (e.g., 1 hour in our case) and for a given technology such as 2G, 3G or 4G~\cite{ibm_nf}. An illustrative example of data feed is shown in Figure~\ref{fig:exp1_labeling}. In our use case, we have access to 2G and 3G data. Tables~\ref{tab:categorical_features} and \ref{tab:network_status_features} present a list of fields captured in the Data Feed. 

\textbf{--~Customer Care Calls}. This dataset is a log of calls received by the telco care center. This collection includes fields such as the date and time of the call, its duration, as well as additional fields identifying the support agent taking the call and his or her level of expertise.

\textbf{Spatial distribution.} Figures~\ref{fig:volume_map}, \ref{fig:map_retransmitted}, and \ref{fig:ccc_map} show the distribution over the geographical area of analysis corresponding to the total download volume in GB, percentage of retransmitted packets, and percentage of customer care callers with respect to the total customers population, respectively. We can observe that the most congested areas, i.e., heavy download traffic, are the ones that suffer a high percentage of retransmissions, which lead to a larger number of customer calls from these areas, a behavior that is intuitively expected. 
%
\begin{figure}[!htb]
\centering
\vspace{2em}
	\includegraphics[width=0.85\linewidth]{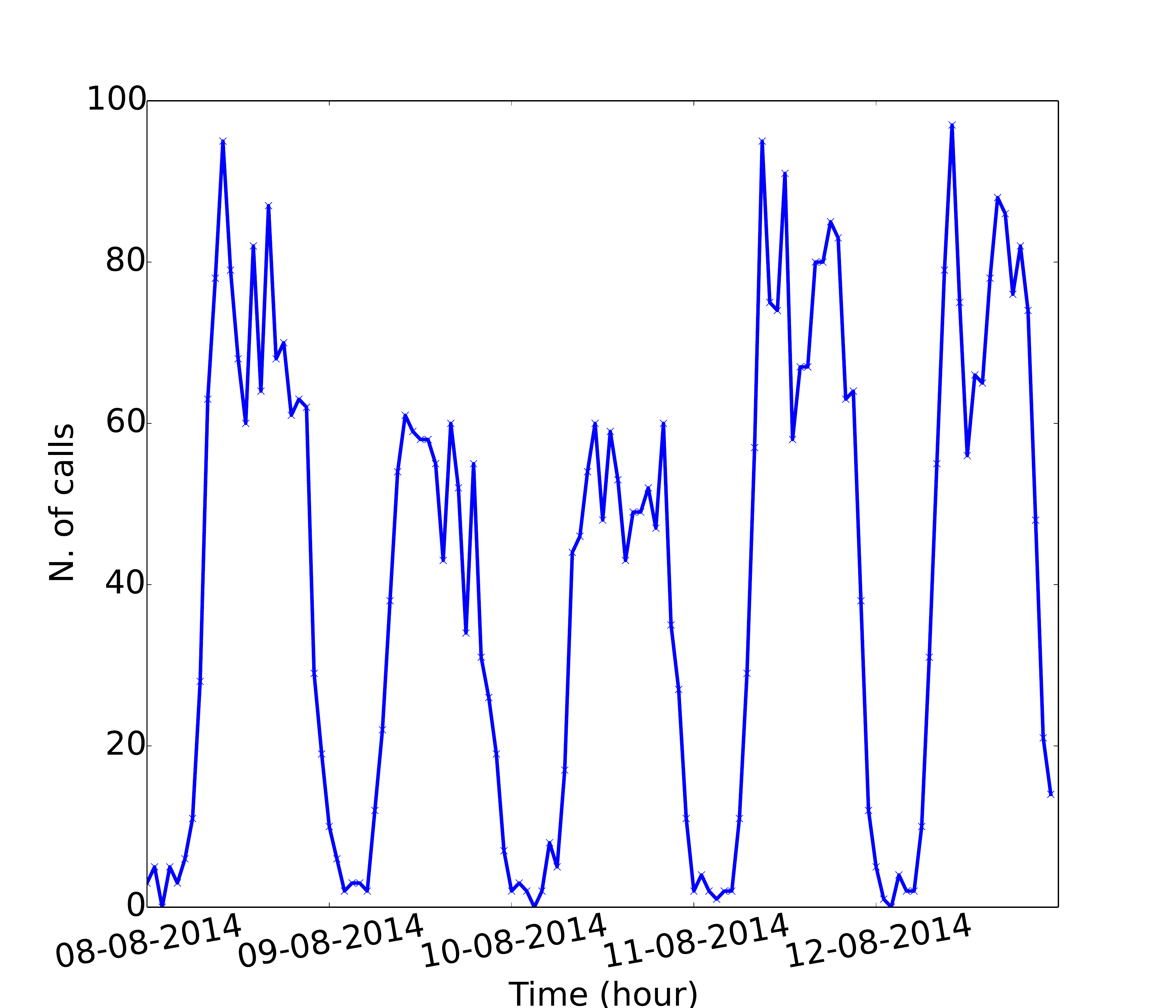}
\caption{Temporal distribution of customer care calls (per hour).}
\label{fig:ccc_hourly}
\vspace{2em}
\end{figure}
\begin{figure}[!t]
\centering
	\subfigure[\tiny Download data (MB) for callers.]{
	\includegraphics[width=0.47\linewidth]{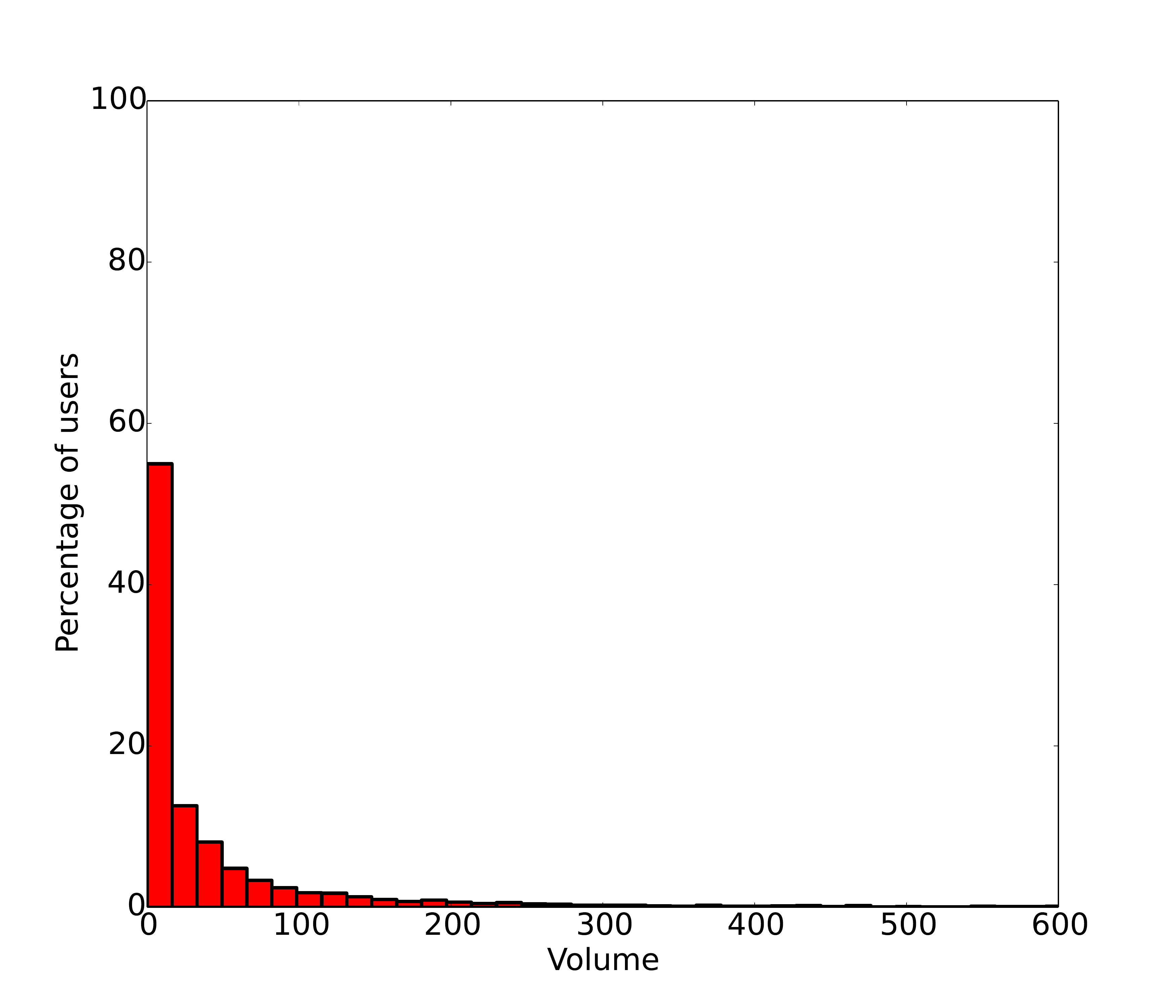}
	\label{fig:subs_callers_volume}
	}~
	\subfigure[\tiny Download data (MB) for no-callers.]{
	\includegraphics[width=0.47\linewidth]{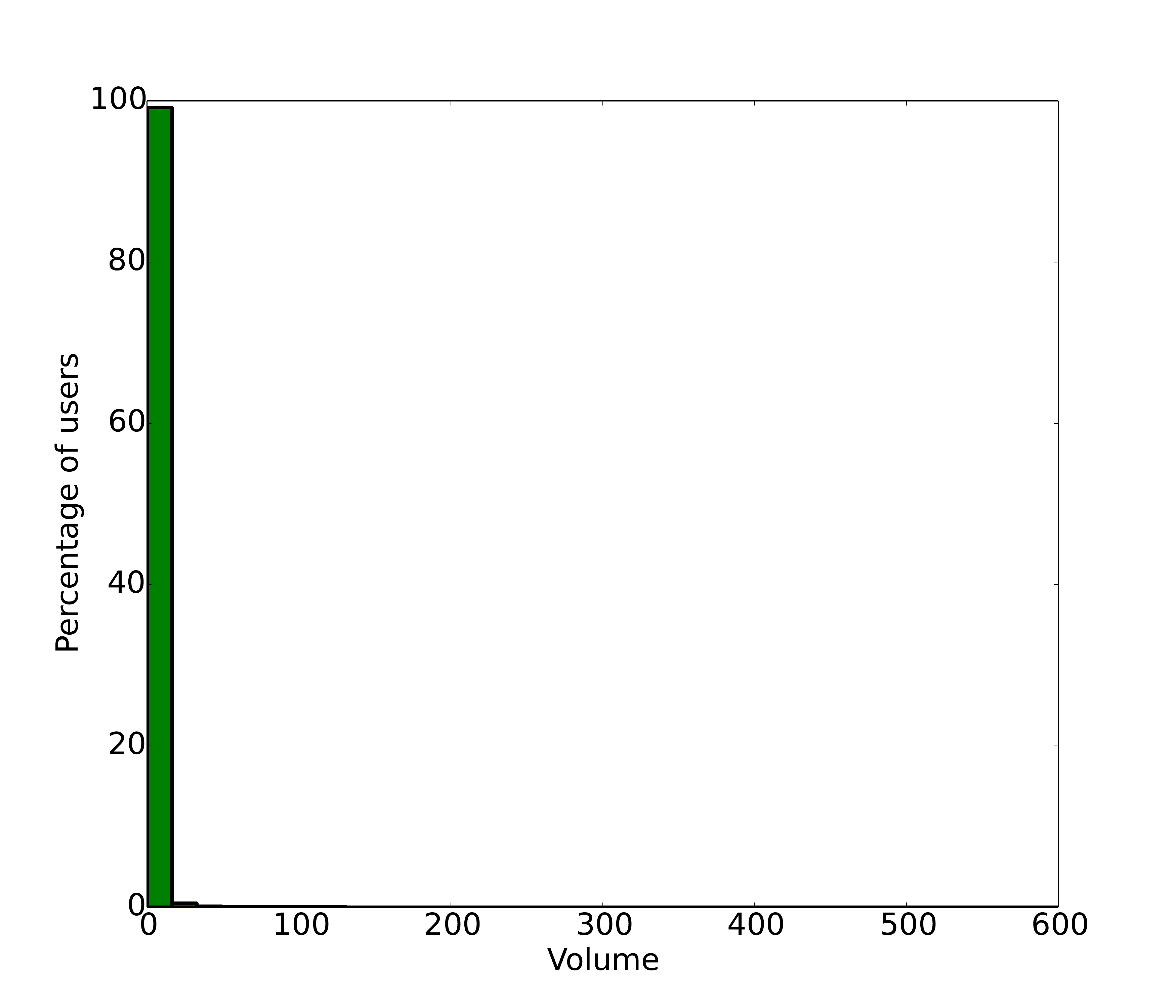}
	\label{fig:subs_no_callers_volume}
	}\\
	\subfigure[\tiny Retransmitted packets for callers (\%).]{
	\includegraphics[width=0.47\linewidth]{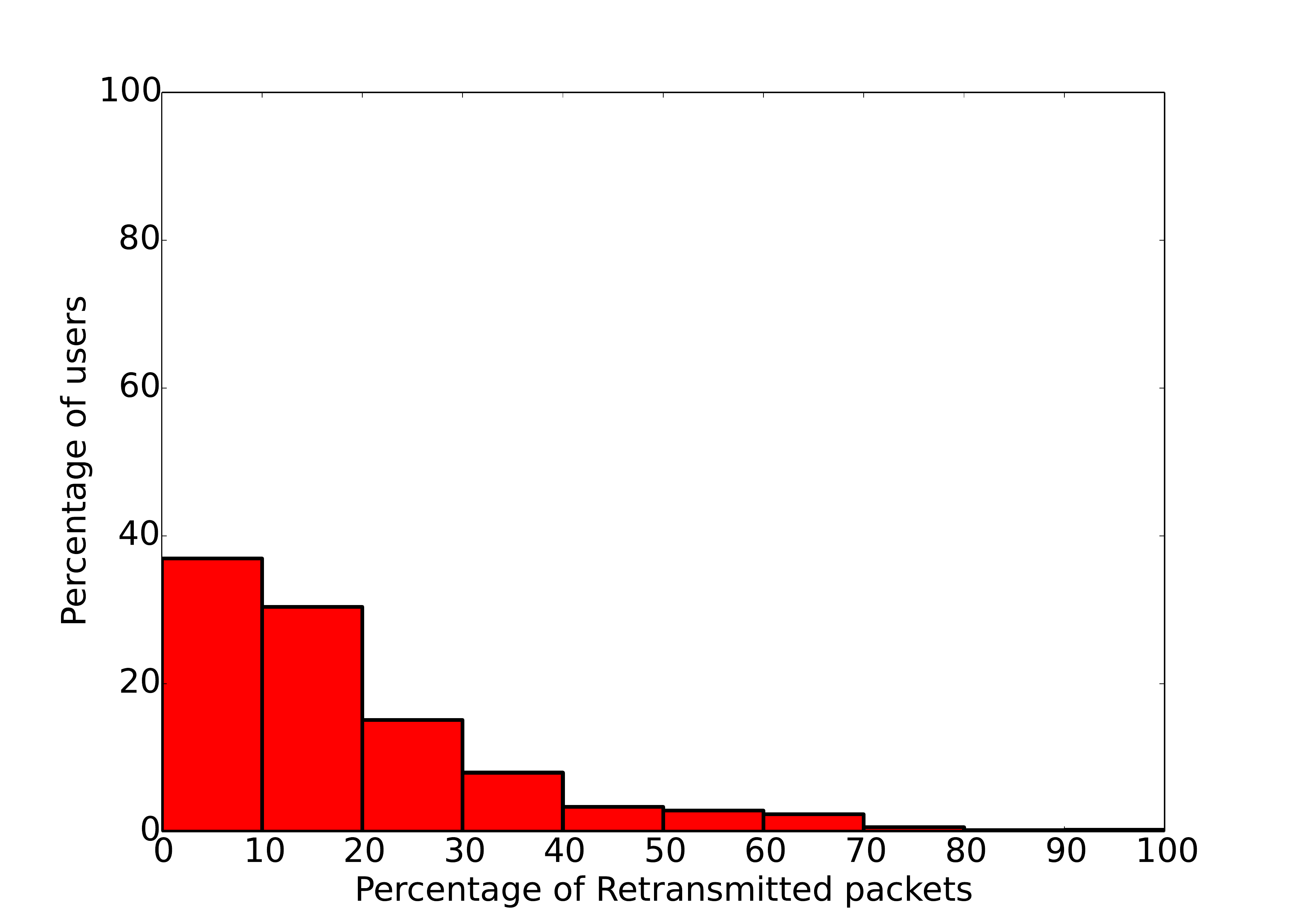}
	\label{fig:subs_callers_retransmitted}
	}~
	\subfigure[\tiny Retransmitted packets for no-callers (\%).]{
	\includegraphics[width=0.47\linewidth]{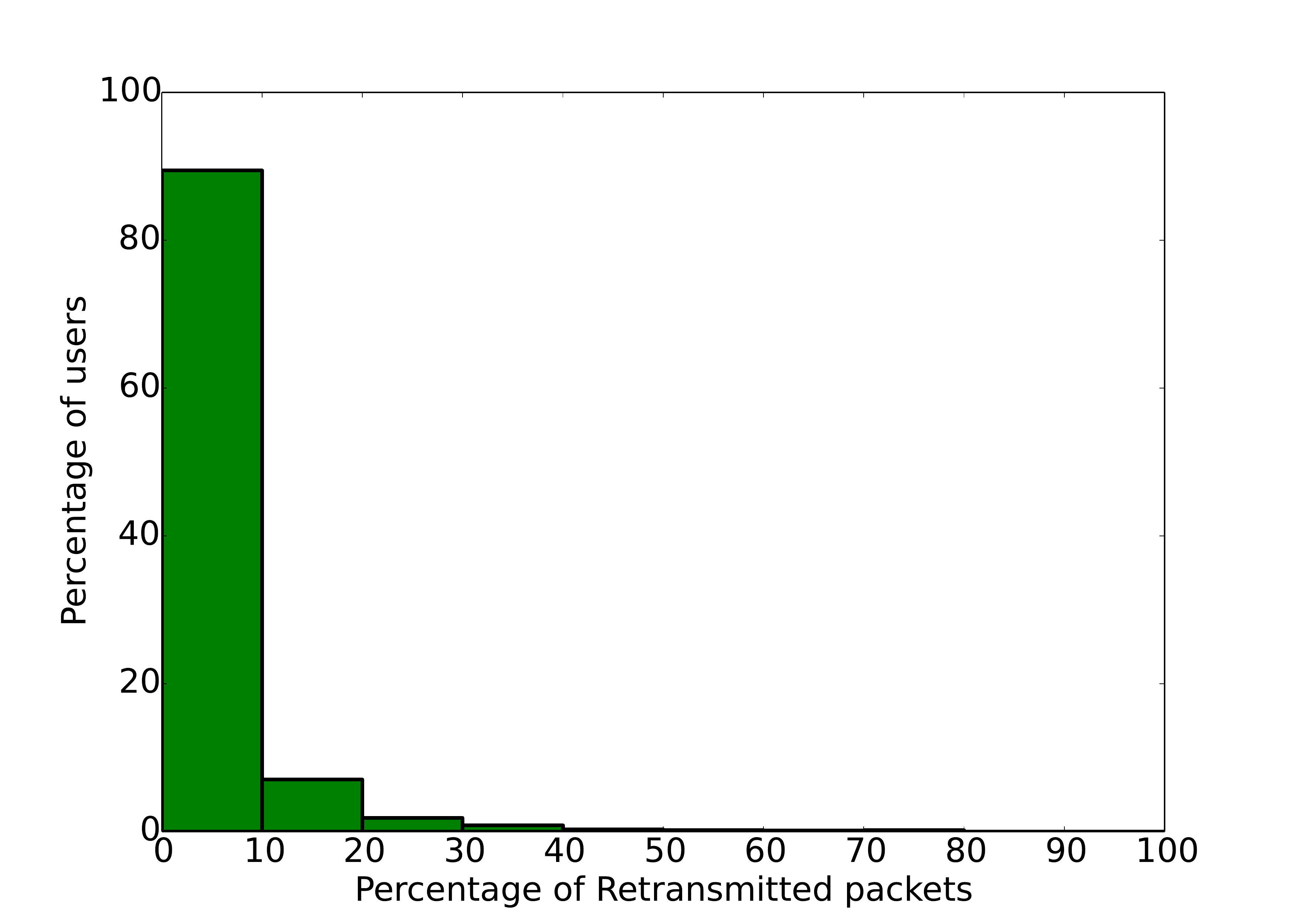}
	\label{fig:subs_no_callers_retransmitted}
	}
\caption{Users that call the care center vs. the ones that do not call.}
\label{fig:dataset_exploration}
\end{figure}
\begin{figure*}[!htb]
\centering
	\subfigure[Top-10 applications.]{
	\includegraphics[width=0.49\linewidth]{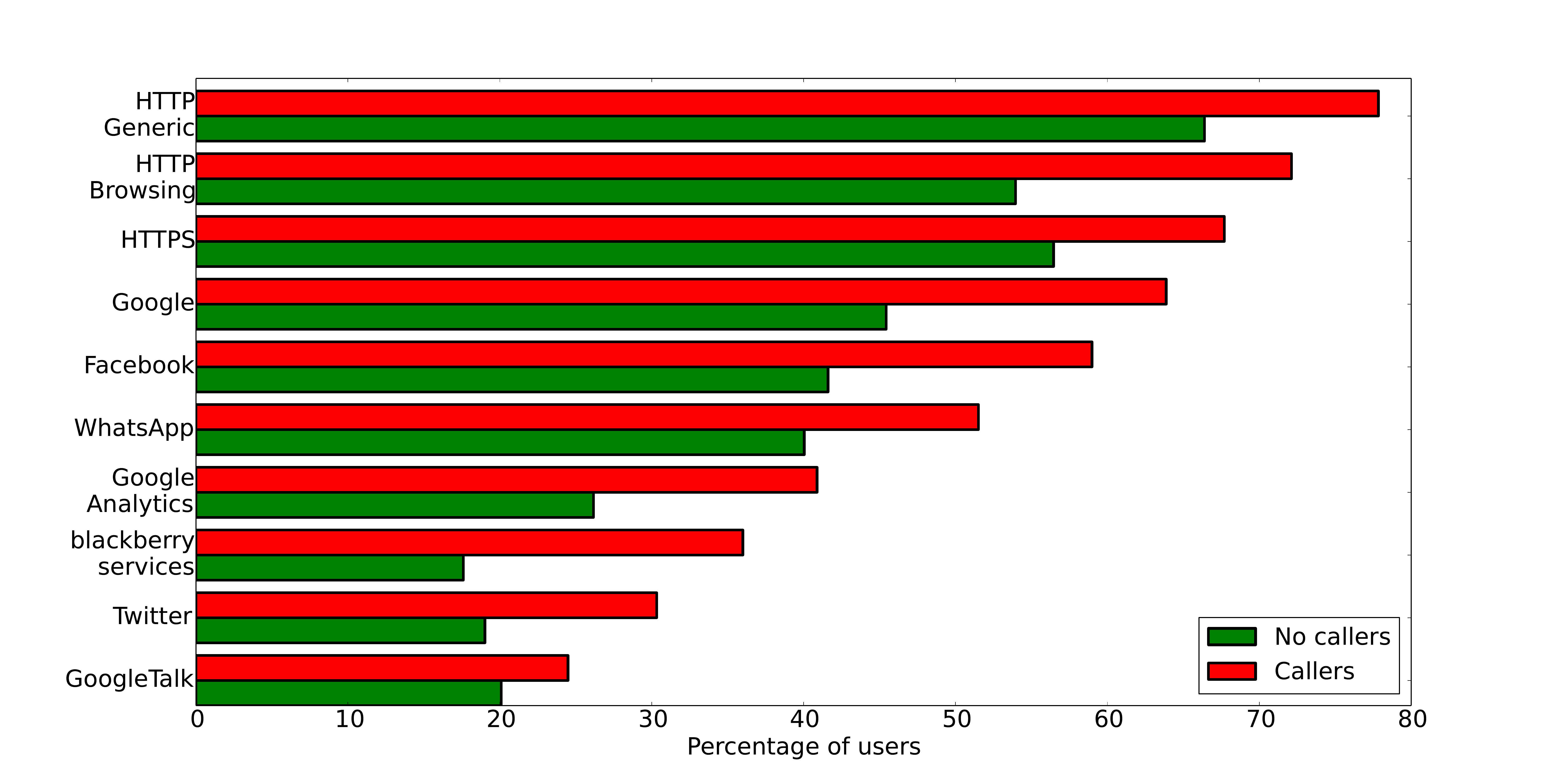}
	\label{fig:subs_apps}
	}~
	\subfigure[Top-10 devices.]{
	\includegraphics[width=0.49\linewidth]{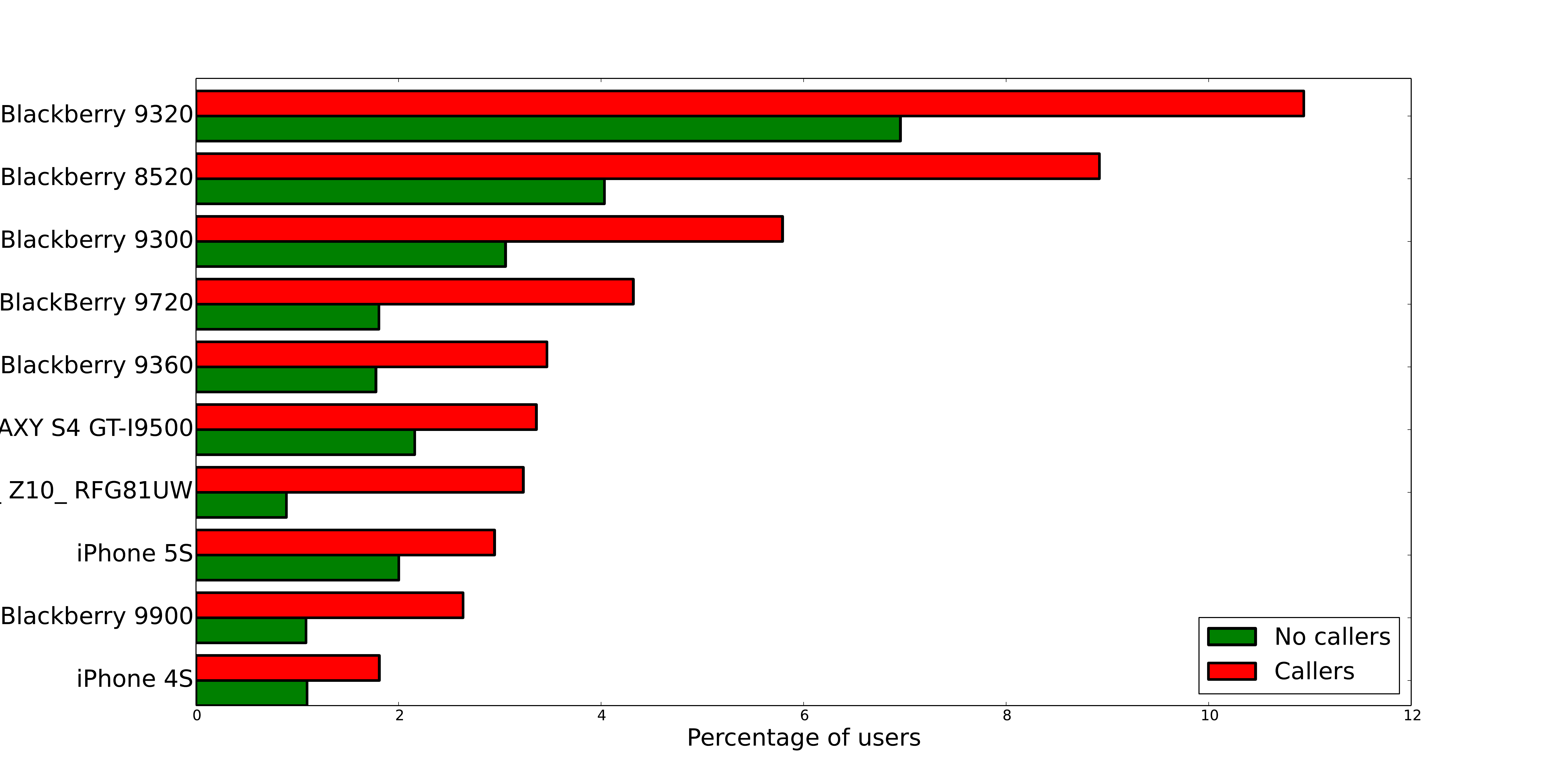}
	\label{fig:subs_model}
	}
\caption{Applications and Devices for \emph{callers} and \emph{no-callers}.}
\end{figure*}

\textbf{Temporal distribution of customer care calls.} In Figure~\ref{fig:ccc_hourly}, we report the hourly number of customer care calls. We can see that the care center receives calls continuously also during the weekend (e.g., 9th and 10th of August).

\textbf{Users that call the care center vs. the ones that do not call.} 
Figures~\ref{fig:subs_callers_volume} and~\ref{fig:subs_no_callers_volume} show the distributions of the total download data volume for users that called the care center (\emph{callers}) and the ones that did not call (\emph{no-callers}), respectively. The callers' distribution has a longer tail to higher download volumes indicating that they are more data intensive users (Figure~\ref{fig:subs_callers_volume}). Similarly, Figures~\ref{fig:subs_callers_retransmitted} and~\ref{fig:subs_no_callers_retransmitted} show the distribution of retransmitted packets for callers and no-callers, respectively. We can highlight again the longer tail on the distribution of the callers indicating that if users experience a higher percentage of retransmitted packets they are more likely to call. Finally, Figures~\ref{fig:subs_apps} and \ref{fig:subs_model} contrast the most popular applications and devices for callers and no-callers.

\hfill

This preliminary analysis shows that user contextual information -- location, type of device, data volume, retransmitted packets, used applications -- can help discriminate users with bad experience. This insight will be used to drive the development of the prediction models in Section~\ref{sec:model}.
%
\section{Our Approach}
\label{sec:model}
Based on our data exploration we observe promising correlations between the data available in the Data Feed and the registered calls to the care center, which can help us with the task at hand of predicting user experience in real-time. To this end, we present in this section the formalization of our approach, which we call Restricted Random Forest (RRF), and discuss the architecture of our deployed proof-of-concept solution.

\subsection{Restricted Random Forest (RRF)}
We cast the problem as a binary prediction task and we use a supervised approach to solve it, while also considering the telco's  requirement to have an explanation of the causes of a bad experience.

Let us denote by $U$ the set of all mobile users in the telecommunications network. For a given time period $B$ and a user $u \in U$, we consider a set of $N$ transactions observed in the data stream of the form $S = \{(x_1,y_1),\ldots,(x_N,y_N)\}_u^B$. Each transaction $x_i$ is represented as a feature vector that summarizes the mobile user context and is associated to a class $y_i$ that takes the value of $+1$ if the customer $u$ places a call to the care center, and $-1$ otherwise.

The idea is to build models offline using data from the recent history, e.g., previous day ($B=1$ day), and once learned the scoring (classification) of the Data Feed transactions is performed online, which allow us to predict if a set of transactions (user experience) will lead the customer to place a call to the care center in the immediate short horizon, e.g., within the current day.

Our RRF solution is based on an ensemble of decision trees~\cite{elemStatLearn} for classification, which are very intuitive and easy to interpret. We learn individual decision trees $T$ to keep interpretability and in order to boost predictive performance we combine them in an ensemble. Let us denote with $M$ the number of decision trees in our ensemble, then the models are combined by majority voting and the most-voted class is predicted:
$$
\hat{y} = f_M(x) = \argmax_y \sum_{m=1}^M I(y = T_m(x; \Theta_m)) \quad ,
$$
where $\Theta_m$ are the model parameters of tree $T_m$ and $I(a)$ is an indicator function that is $1$ if $a$ is $true$ and 0 otherwise.

Our approach is clearly inspired by random forest ensembles~\cite{elemStatLearn}, the main difference is that we explicitly specify the features to be used in each training bag, rather than using a random subset of features as in random forest.

Remember that one of the main requirements from the telco for this solution was to emphasize interpretability, even at the expense of predictive performance.

\subsection{Architecture}
\label{sec:architecture}
The deployment of our solution requires a system with two key capabilities:

\noindent\emph{(1)}~The ability to process massive amounts (in the order of petabytes) of historical telco network and call center data in order to build predictive models based on previous customer behaviour.

\noindent\emph{(2)}~The ability to process fast high bandwidth telco network data (in the order of terabytes per hour) for predicting customer behavior based on recent experience and historical models.

\begin{figure*}[!tb]
\centering
	\includegraphics[width=0.8\linewidth]{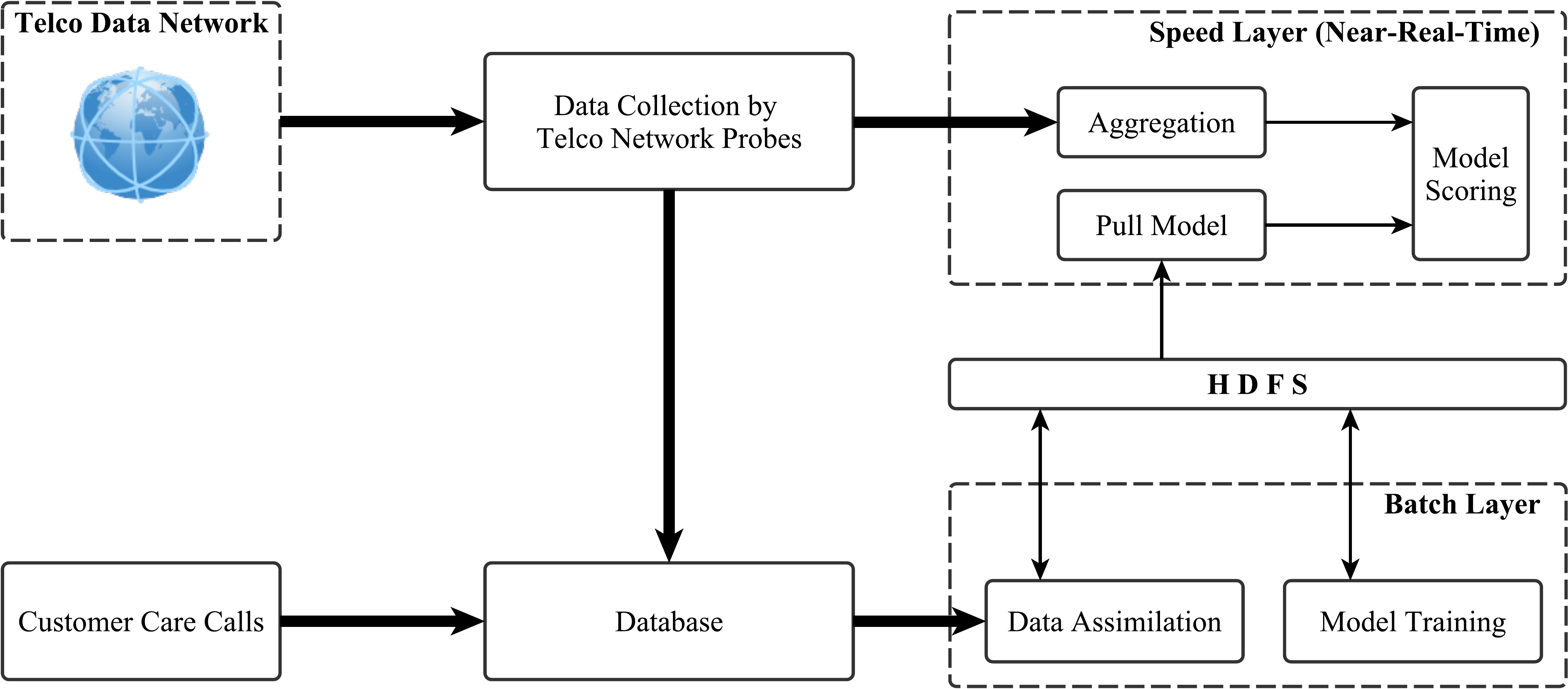}
\caption{System architecture for modeling and scoring telco network data.}
\label{fig:system_architecture}
\end{figure*}

In other words, our architecture requires a \emph{batch layer} for model training and a \emph{speed layer} for model scoring.  In practice, the scale of the historical data prohibits the learning of predictive models in a timely manner without employing a parallel approach. The greater relevance of recent telco and call center data to the predictive models makes having short batch cycle times even more critical (i.e., the longer the batch cycle the more outdated current models are by the end of the cycle).  Thus we partition the telco data by date, and train independently on each partition, as reflected in our evaluation (Section~\ref{sec:experiments}).  This approach to partitioning simplifies data ingestion and is effective when deploying models as it facilitates the weighting of models built using newer historical data, over those built using older historical data, if needed. Figure~\ref{fig:system_architecture} sketches the architecture of our solution.
%

It is instructive to compare the described system architecture with systems adhering to the Lambda architecture model for data stores~\cite{marz2015big}. Such systems consist of three main components --~a batch layer, a speed layer, and a serving layer~-- and like our proposed system input data is fed to both the batch layer and the speed layer. The proposed system shares many aspects of a Lambda architecture based system, namely the speed layer, batch layer, and simultaneous ingestion. The goals of our system differ to those of a data store though, that is, our system currently has no requirement for a serving layer for unified querying of both the batch and speed layers. Rather the batch layer exports a model directly to the speed layer, which can be used to trigger actions based on real-time predictions. 

\subsubsection*{Infrastructure of our Solution.}
We now describe in more detail the subset of the overall system used in our on-site evaluation at the telco. This corresponds to the batch and speed components shown in Figure~\ref{fig:system_architecture}. We used the \emph{IBM BigInsights for Apache Hadoop} product~\cite{2015:ibm:bi} for the batch processing middleware layer and the \emph{IBM Infosphere Streams} product for the speed processing middleware layer~\cite{2015:ibm:streams}.

BigInsights was installed on a \mbox{4-node} cluster of IBM Power systems, and Infosphere Streams was installed on a separate IBM Power system machine. The historical telco data was stored in HDFS\footnote{HDFS: Hadoop Distributed File System} on the BigInsights cluster and a component of BigInsights called \emph{Big R} (a middleware layer for distributing \emph{R} code) was used for off-line model building.  Infosphere Streams was used for real-time scoring of live telco data\footnote{For the purposes of our evaluation we read the stored telco data from disk rather than live from the telco network.} by deploying the models built using Big~R.  BigInsights models were shared with Infosphere Streams by generating a PMML\footnote{PMML: Predictive Model Markup Language} version of the model and persisting it within HDFS.

\section{Empirical Evaluation}
\label{sec:experiments}
As discussed earlier, it is very difficult to determine what is a \emph{good} or \emph{bad} user experience at every single moment of the day and without asking the users for direct feedback. What then is the best way to infer bad experience from a time series of Internet usage information? To answer this question we test two approaches:

\noindent\textbf{--~Approach~I.} Here we use the observed user transactions in the Data Feed before the user places a call to the care center as proxy for bad experience. We also (optimistically) assume that the issue motivating the call is solved by the care center, so the following user activities reflect a good experience. Thus we label each user transaction accordingly and train our model ensemble. For real-time classification, we used the ensemble to score each new transaction from the Data Feed and then aggregate, for each user, her transactions and predicted class (will call the service center or not) in order to assess what is the overall user experience during a given time period (e.g., 1 day).

\noindent\textbf{--~Approach~II.} For this approach, we assume that the issue motivating the call is the result of a perceived poor experience in a relatively long time period. Thus, differently from Approach I, we aggregate and summarize the transactions per user over a period of time \emph{before} training the classifiers.
%

We conduct two experiments in this section in order to evaluate each of the aforementioned approaches. We first detail the experimental setting common to both assessments and then report the results for Experiment~I and Experiment~II, which respectively explore approaches~I and II. Finally, we conclude this section with a discussion of our findings.

\subsection{Experimental Settings}
We partition the telco data using a daily temporal split. We reserve as the test set all Data Feed transactions from the last day in our collection, i.e., 2014-08-12, and the rest of the data constitutes the training set. 
\\

We use precision, recall, and f1 as the metrics to assess the performance of our approach. They are defined as follows. 

$$
\text{precision} = \frac{TP}{TP + FP}
$$
$$
\text{recall}    = \frac{TP}{TP + FN}
$$
$$
\text{f1}        = 2 * \frac{\text{precision} * \text{recall}}{\text{precision} + \text{recall}}
$$
where TP, FP, and FN are the number of true positives, false positives, and false negatives, respectively.

We prefer these metrics over accuracy given the unbalanced nature of the problem. Note that, in our collection, only a small percentage of all users made a call to the service center. Thus, obtaining a high accuracy would be trivial, e.g., by predicting no calls at all to the care center.

For the experiments, we build the decision trees, which are the individual members of our RRF ensemble, in a distributed fashion using Big~R~(cf. Section~\ref{sec:architecture}). We build each decision tree model using bootstrap aggregating or bagging~(e.g., \cite{ensembleLearning:polikar:2006,ensembleLearning:2009}), to this end we use the Big~R's \texttt{bigr.frame}~\cite{2015:ibm:bigr} for Hadoop / MapReduce and the \texttt{rpart} R package~\cite{2014:rpart}, we experimentally set the minimum split parameter equal to 10 observations, the maximum depth of the tree to 10, and the complexity parameter (cp) to 0.001 (used by default in our experiments if not specified otherwise).

In order to deal with the class imbalance of our dataset, the bagging procedure includes in each \emph{training bag} all $n^+$ positive examples available in the original training set and a random sample, with replacement, of size $n_{bag}^-$ of the original negative examples. We experimented with different sizes for  $n_{bag}^-$ and found that a fully balanced setting ($n_{bag}^- = n^+$) led to good results.

In the experiments we compare our RRF approach against the following state-of-the-art classifiers using the same data representation:
\begin{itemize}
\item Support Vector Machine (SVM)
\item Logistic Regression (LR)
\item Gradient Boosted Trees (GBT)
\item Random Forest (RF)
\end{itemize}


\subsection{Experiment I}
\label{sec:experiment_1}
%
In Experiment~I we explore a data representation based directly on the transactions captured from the data feed in order to assess Approach~I, which was introduced earlier in this section. Each transaction corresponds to an aggregation, per hour, of different activities for each user (e.g., browsing, video streaming). 

In this experiment we focus on a 24-hour time span, where the user experience for this period is given by the set of her transactions within the given day. We use the observed user transactions before she placed a call to the care center as proxy for bad. Furthermore, we optimistically assume that the issue leading to the make the call is solved by the care center, so the following user activities (i.e., transactions) reflect a good experience. This approach allows us to label our training feature vectors.
%
%
%

For example, consider the user \emph{Alice} whose transactions are illustrated in Figure~\ref{fig:exp1_labeling}. In this particular day, Alice placed a call to the care center at 14:30 hours complaining about her bad experience, therefore, all Alice's transactions before 14:30 represent the bad experience causing Alice to call and complain to the care center. The agents at the service center solved Alice's problem and all following transactions (after 14:30) are considered a good experience for Alice.

Note that for users who made more than one call to the care center during a given day, we consider only the latest call placed. Thus, all previous transactions to the last call correspond to a bad experience and the following ones after the call as good experience, e.g., issues that caused the complaint were not solved in earlier calls to the care center.

\begin{table*}[!ht]
\caption{Categorical Features Extracted from Data Feed. GPRS refers to the General packet radio service, which is a packet oriented mobile data service on the 2G and 3G cellular communication system's global system for mobile communications (GSM).}
\label{tab:categorical_features}
\centering
\setlength{\tabcolsep}{3pt}
\scalebox{1}{
\begin{tabular}{ l  p{39em} }
\toprule
\textbf{Feature} & \textbf{Description} \\
\midrule
Protocol & Protocol used during the transaction. Values: tcp and udp. \\
\midrule
Application & Application used during the transaction -- Android Market, Apple Maps, WhatsApp, etc. \\
\midrule
APN & Access Point Name. Examples: internet, mobi, lte, etc. \\
\midrule
SGSN & In a GPRS network, SGSN handles all packet switched data within the network.\\
\midrule
GGSN & GGSN internetworks between the GPRS and external packet switched networks.\\
\midrule
Cell & Cell associated to the transaction.\\
\midrule
Location Area & location area of the transaction.\\
\midrule
Device Manufacturer & E.g.: HTC, Huawei, LG, Nokia, Samsung, Sony-Ericsson.\\
\midrule
Device Model & E.g.: HTC One, Huawei Ascend, Nexus 5, Nokia Lumia, Samsung Galaxy \\ 
\midrule
QoS & Quality of service category. Values: Gold, Silver, and Bronze. \\
\midrule
RAT & Radio Access Type. Values: GERAN and UTRAN.  \\
\midrule
Rule type & Aggregation dimension for customer summary: video or web browsing.\\
\bottomrule
\end{tabular}
}
\vspace{5em}
\end{table*}

\begin{table*}[!ht]
\caption{Network Status Features Extracted from Data Feed.}
\label{tab:network_status_features}
\centering
\scalebox{1}{
\begin{tabular}{| l | p{20em} | l | p{32em} |}
 \hline
\textbf{\#} & \textbf{Description} & \textbf{\#} & \textbf{Description} \\ \hline
1	&	Bytes Up	& 28	&	Minimum download TCP User Throughput	\\ \hline
2	&	Bytes Down	& 29	&	Upload Application User Throughput: Uploaded Bytes in each range	\\ \hline
3	&	Bytes Total	& 30	&	Upload Application User Throughput: Time Spent in each range	\\ \hline
4	&	Uplink Packets	& 31	&	Download Application User Throughput: Downloaded Bytes in each range	\\ \hline
5	&	Downlink Packets & 32	&	Download Application User Throughput: Time Spent in each range	\\ \hline
6	&	Total Packets & 33	&	Total Round Trip Time	\\ \hline
7	&	Retransmissions Up Link	 & 34	&	Count of Round Trip Time	\\ \hline
8	&	Retransmissions Down Link & 35	&	Total Round Trip Time (Smoothed)	\\ \hline
9	&	Retransmissions Total & 36	&	Count of Round Trip Time (Smoothed)	\\ \hline
10	&	Server Setup Time & 37	&	TCP RTT Distribution: Packets in each range	\\ \hline
11	&	Count of Server Setup Time & 38	&	Maximum upload Application User Throughput	\\ \hline
12	&	Client Setup Time & 39	&	Application User Throughput Bytes Up	\\ \hline
13	&	Count of Client Setup Time & 40	&	Application User Throughput Active Time Up	\\ \hline
14	&	Time to First Byte & 41	&	Maximum upload User Throughput	\\ \hline
15	&	Count of Time to First Byte & 42	&	User Throughput Bytes Up	\\ \hline
16	&	Download Time & 43	&	User Throughput Active Time Up	\\ \hline
17	&	Count of Download Time & 44	&	Maximum download Application User Throughput	\\ \hline
18	&	Upload Time & 45	&	Application User Throughput Bytes Down	\\ \hline
19	&	Count of Upload Time & 46	&	Application User Throughput Active Time Down	\\ \hline
20	&	New TCP Connection Count & 47	&	Maximum download User Throughput	\\ \hline
21	&	Throughput Bytes Up & 48	&	User Throughput Bytes Down	\\ \hline
22	&	Throughput Upload Time & 49	&	User Throughput Active Time Down	\\ \hline
23	&	Throughput Bytes Down & 50	&	Video Time to Start Distribution: Video Count in each range	\\ \hline
24	&	Throughput Download Time & 51	&	Video Playback Gap Ratio Distribution: Total Play Time in each range	\\ \hline
25	&	Maximum upload TCP User Throughput & 52	&	Video Bitrate Distribution: Playback Time per Range	\\ \hline
26	&	Minimum upload TCP User Throughput & 53	&	Volume of Video downloaded: Bytes per Range	\\ \hline
27	&	Maximum download TCP User Throughput	& 54	&	Time to First page object displayed: Count in each range	\\ \hline
{--} & {--} & 55 &	Full web page load time: Count in each range \\ \hline
\end{tabular}
}
\vspace{5em}
\end{table*}

\noindent\textbf{Data Representation and Models.} We build feature vectors based on the data feed transactions. We represent categorical features (cf. Table~\ref{tab:categorical_features}) using one-hot encoding, that is, binary vectors with a single dimension of the vector equal to 1 denoting the category, and 0 otherwise. For example, in the dataset there are 178 applications (e.g., blackberry\_services, dns, https, facebook), thus the binary vector representing the feature \emph{application} is of size 178. For example, if the application `facebook' is 5th in the list, its binary categorical representation will be a vector of 178 dimensions with all zeros, except for the one in the 5th position. A similar approach is taken for the rest of categorical features.

For the variable `cell' that has a large number of categories (more than 13K different values), we opted to represent the 16 more common cells, in order to reduce sparseness while still capturing a large part of the user activities within the cells with high traffic. A similar approach was taken to represent the `device models', using only the 16 most popular devices. In these two cases, an additional bit was added to the representation, which is activated (i.e., taking value of one) if any of the other values outside the top 16 was observed.

Real-valued features denoting the network status, e.g., bytes up and down, download time, retransmitted packets, etc., are represented directly with their corresponding numerical value extracted from the data feed. In Table~\ref{tab:network_status_features} we provide the complete list of the network status features. In our models we use feature vectors of size 480, which include all categorical and real-valued features to represent the user context.

We learn decision trees from selected variables of interest for the telecom operator, which will help them to understand the results. 

In particular, we use the following variables to build the members of our ensemble: 
\begin{enumerate}
\item Application
\item Application\_Type
\item Cell
\item Location Area
\item Device Model
\item Device Manufacturer
\item Network KPIs
\end{enumerate}

For the models corresponding to the location area we used a complexity parameter (cp) value of cp=0.00015. For the rest we used the default setting for our experiments, i.e., cp=0.001.

For all classifiers, we set their parameters via cross validation, and the corresponding values are as follows. We trained the SVM for 100 epochs using Stochastic Gradient Descent (SGD) with an initial learning rate $\eta_0=0.1$ and decreased it gradually at every iteration, the regularization constant was set to $\lambda = 0.0001$. The Logistic regression model was regularized using a complexity constant $C=1$. For the GBT and RF ensembles, we used 250 and 2000 estimators, respectively, with a maximum depth of 4 for each of decision trees members of the ensemble. We set the maximum number of features to consider when splitting a node equal to the square root of the number of features in our data. Furthermore, for GBT we set a learning rate of 0.1.
%

\begin{figure*}[htb]
\centering
\subfigure[Experiment~I]{
	\includegraphics[width=0.45\textwidth]{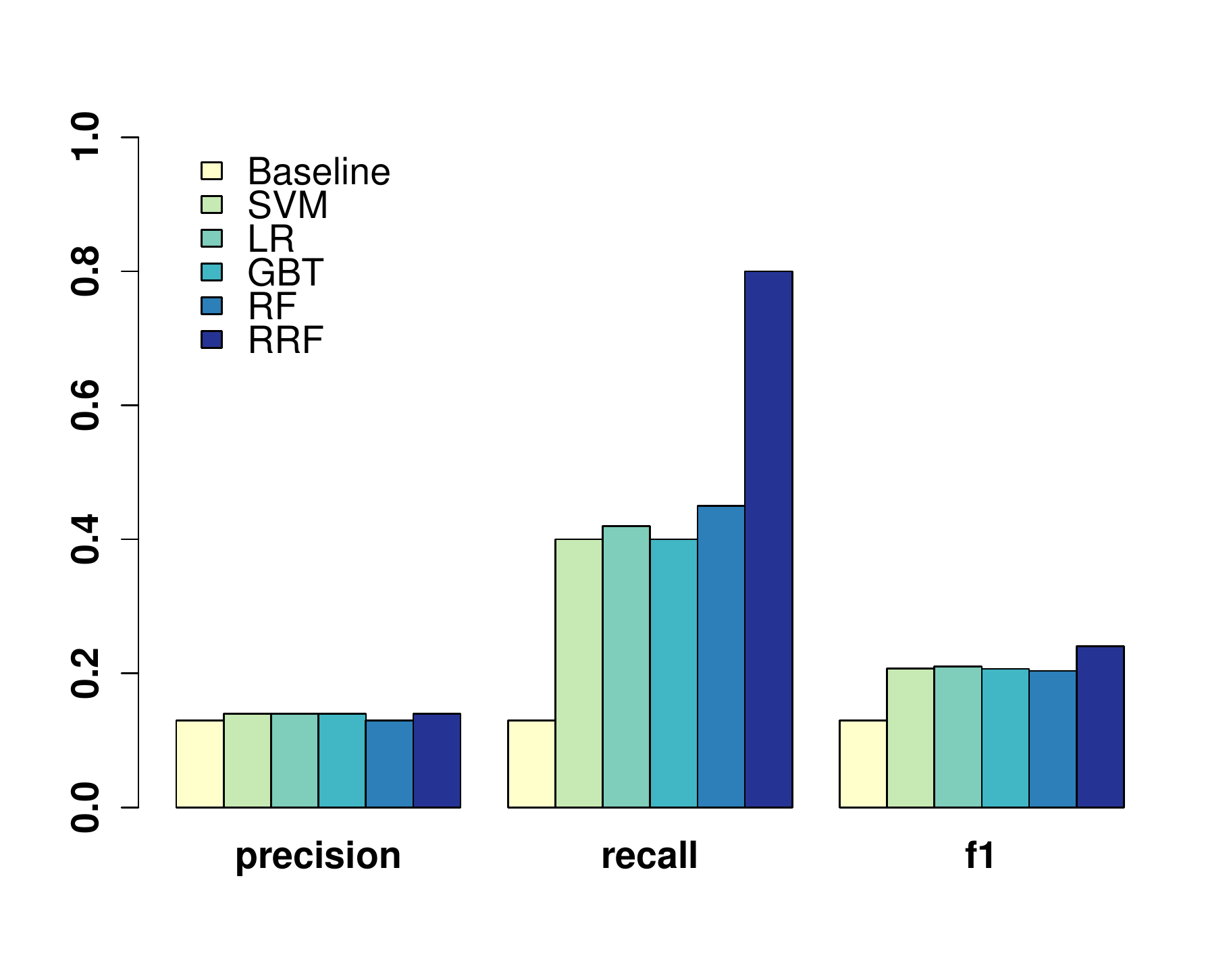}
	\label{fig:exp1_performance_user_aggregate}
}\qquad
\subfigure[Experiment~II]{
	\includegraphics[width=0.45\textwidth]{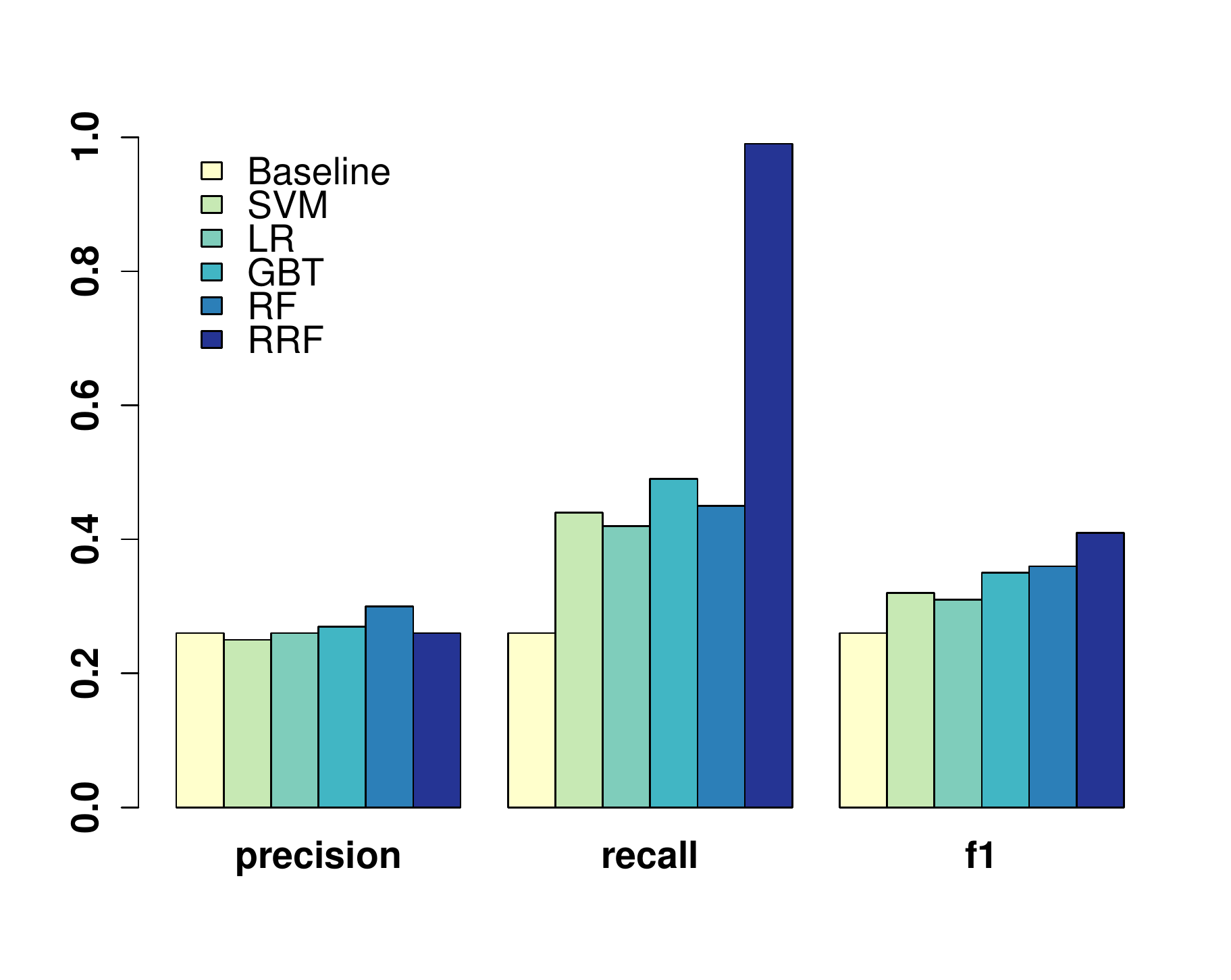}
	\label{fig:exp2_performance_user_aggregate}
}
  \caption{User-based predictive performance.}
  \label{fig:exp1_and_exp2_performance_user_aggregate}
\end{figure*}

\noindent\textbf{Results.} 
The performance for the prediction of the overall user experience in the testing day is shown in Figure~\ref{fig:exp1_performance_user_aggregate}. In order to compute this per user aggregate, for all models as well as for the baseline,  we use majority voting on the user transactions in the test set. That is, if the user has more transactions predicted as ``call" than the ones predicted as ``no-call", then the overall user experience for the given day is considered a bad one, and a good experience otherwise.

The baseline corresponds to a model that predicts for the test day (2014-08-12) based on the transactions distribution of call/no-call from the previous day (2014-08-11). More precisely, for each transaction in the test set, we predict if it will lead to a call or not to the care center by tossing a biased coin with 13\% probability of labeling as positive (call) and 0.87\% otherwise. 

Table~\ref{tab:exp1:variable_importance} lists the most important variables reported by the \texttt{T\_Application}, \texttt{T\_Device\_Model}, and \texttt{T\_Network\_KPI} decision trees, built from the Application, Device Model, and Network KPIs variables, respectively.

\begin{table}[!t]
\caption{Most important variables for classification\break according to the decision trees.}
\label{tab:exp1:variable_importance}
\centering
\setlength{\tabcolsep}{3pt}
\scalebox{1}{
\begin{tabular}{l p{20em}}
\toprule
\textbf{Model} & \textbf{Most Important Variables} \\
\midrule
\texttt{T\_Application}   & BlackBerry services, Facebook, http (generic), Twitter, Tango, BlackBerry messenger, Android market, Apple Maps, Microsoft Push Notification Service (MPNS), MSN Webmail.\\
&\\
\texttt{T\_Device\_Model} & BlackBerry 9720, iPhone 4s, BlackBerry 9300, BlackBerry 9360, iPhone5. \\
&\\
\texttt{T\_Network\_KPI}  & Client/Server setup time, packets count (up, down, total), bytes count (up, down, total), upload time, new TCP connections,  retransmitted packets (up, down, total), application round trip time, TCP round trip time.  \\
\bottomrule
\end{tabular}
}
\end{table}
%
%
%

\subsection{Experiment II}
\label{sec:experiment_2}
Here we explore a different approach to characterize the bad experiences of users w.r.t. the approach described in Experiment~I.

We assume that the level of service was poor in the latest 24 hours in which the user made a call. Instead of focusing on each single transaction, we provide a summary per application of the user experience in the last 24 hours.

For example, considering Alice's (unlabeled) transactions in Figure~\ref{fig:exp1_labeling}, the corresponding summaries for YouTube and AppleMaps, could look as follows, both labeled as bad experience, since Alice called the service center that day:
\begin{center}
\vspace{0.5em}
\includegraphics[width=0.95\columnwidth]{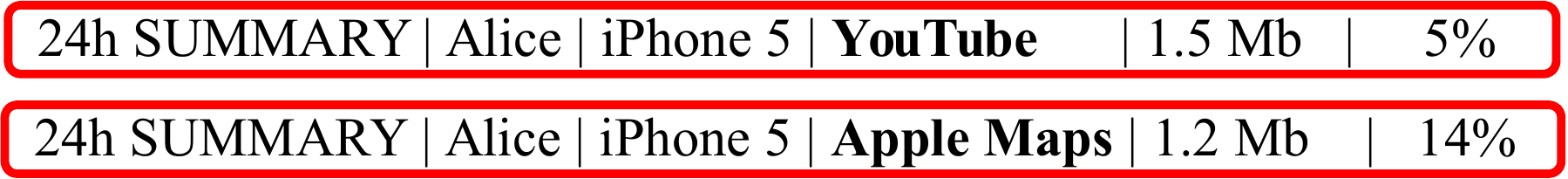}
\vspace{0.5em}
\end{center}


\begin{table}[!t]
\caption{Features per application for Experiment~II.}
\label{tab:feature_exp2}
\centering
\setlength{\tabcolsep}{5pt}
\scalebox{1}{
\begin{tabular}{l p{25em}}
\toprule
\textbf{} & \textbf{Feature} \\
\midrule
1 & Total bytes downloaded in the last 24 hours. \\
\midrule
2 & Hourly average number of retransmitted packets. \\
\midrule
3 & Max. time needed by the user to receive the first byte from an application in the last 24 hours. \\
\midrule
4 & Min. download time experienced by the user in the last 24 hours. \\
\midrule
5 & Min. upload time experienced by the user in the last 24 hours. \\
\midrule
6 & Min. hourly averaged Round Trip Time in the last 24 hours.\\
\midrule
7 & Min. hourly averaged upload throughput. \\
\midrule
8 & Min. hourly averaged download throughput.\\
\midrule
9  & Hourly averaged customer experience score (CES), a metric heuristically computed by the telco based on the network status.\\
\bottomrule
\end{tabular}
}
\end{table}

\noindent\textbf{Data Representation and Models.}
We build the summaries of user mobile activity considering the most visited cell, the device used, and aggregated network-related features or key performance indicators (KPIs) for the top 10 applications in terms of downloaded bytes across all the users. 

We decided to consider only those applications since they represent the ones with more interactions with the network and the ones that might have higher expectations for the users. Therefore if a poor Quality-of-Service affects these applications is more likely that a user complains about the network in general.

The top 10 applications are: \emph{HTTP}, \emph{BlackBerry services}, \emph{Facebook}, \emph{Google}, \emph{Apple Maps}, \emph{YouTube}, \emph{Android Market}, \emph{SSLv3}, \emph{Twitter}, \emph{WhatsApp}, which correspond to the dimensions for the feature vectors for Experiment~II.

For each of these applications we use the features described in Table~\ref{tab:feature_exp2} to compute the aggregated KPIs.

Regarding the most visited cell for a subscriber, this represents the cell where the user spent most ``active'' time. We do not report all the cells, but we label as ``other cell'' the set of cells with less than 7 users, since the mean for the antenna distribution is 6.8. We consider 751 distinct locations in total.

A similar approach has been used for the \emph{Device Model} feature, where we use a dummy field, ``other device'', for all devices used by less than 5 users, which corresponds to the median of the device distribution. We consider 193 different devices. We use the median instead of the mean to select the cut-off point in this case, because the device distribution is much more skewed than the cell's one.

For instance, considering the Apple Maps application, we aggregate all the downloaded bytes for a specific user in the last 24 hours. Note that this feature set takes into account an heuristic computed by the telco, named Customer Experience Score -- CES, that corresponds to predefined quality thresholds of user network variables. For instance, if the minimum download throughput is high, the CES is consider to be \emph{poor}.

In total, for each user we aggregate her experience at the daily level, and we label this summary as positive or negative based on whether she made a customer care call in that day or not.

In Experiment~II we trained the SVM for 1000 epochs also using SGD. The initial learning rate was set to $\eta_0=0.001$ and the regularization constant to $\lambda = 0.01$. For the rest of the classifiers, we found that the same parametrization used for Experiment~I yielded the best results during cross validation.
%
%
%
%
%

\noindent\textbf{Results.} The baseline is computed as follows. We consider the ratio of positive and negative samples in the day before the evaluation (2014-08-11). For each user in the test day (2014-08-12), we predict based on this ratio whether she will call or not the care center by tossing a biased coin with 25.77\% probability of labeling as positive (call) and 74.23\% labeling as negative (no call). 

The results are shown in Figure~\ref{fig:exp2_performance_user_aggregate}. Considering the f1 score, our RRF approach achieves the best f1 score, our ensemble improves roughly 59\% the results obtained with the baseline model. Random Forest (RF) delivers the best precision against all models at the expense of recall.

\begin{figure}[!t]
\centering
\includegraphics[width=0.95\linewidth]{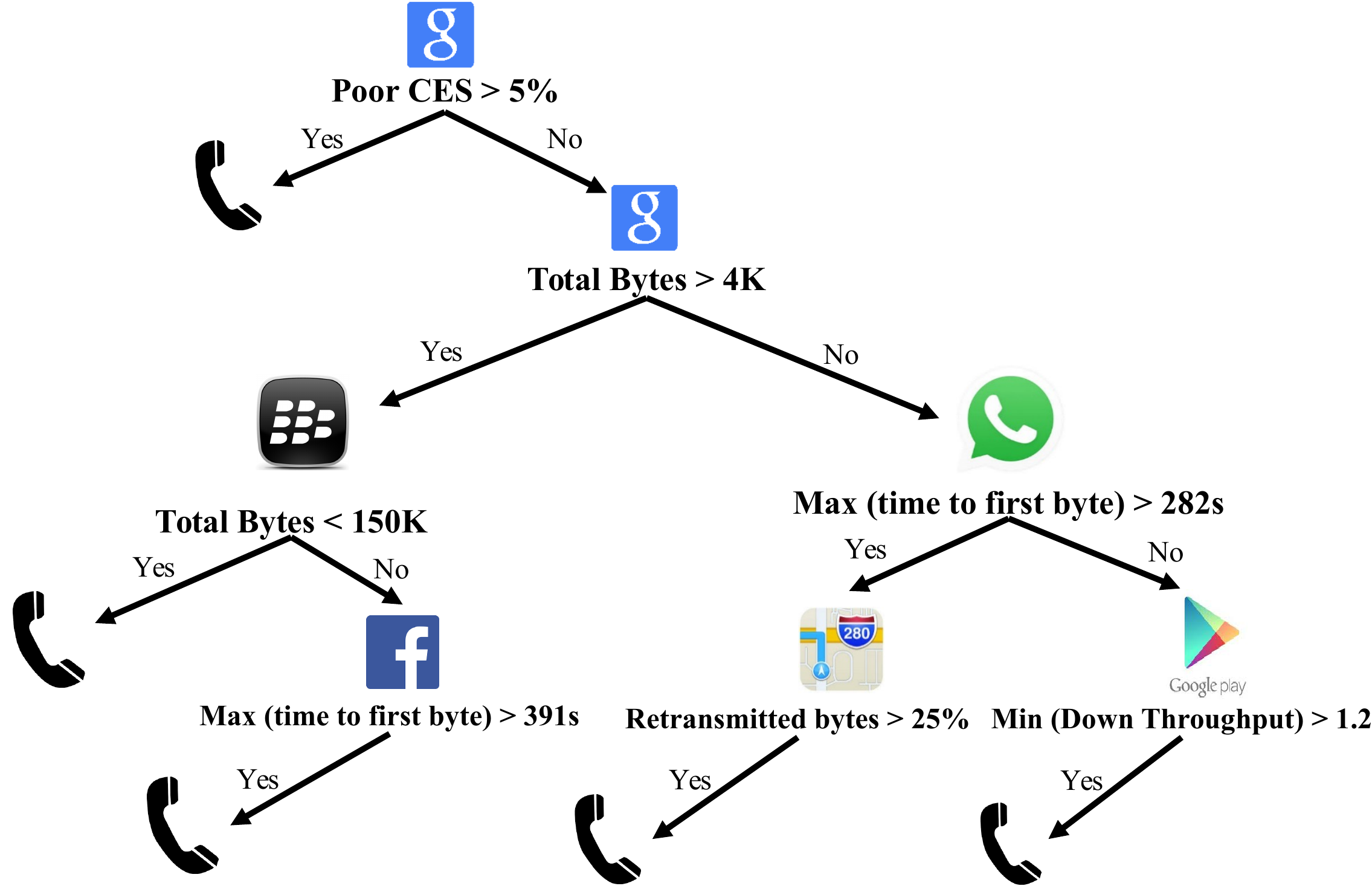}
\caption{Model built using aggregated KPIs computed for a subset of the applications.}
\label{fig:trees_experiment2}
\end{figure}

\subsection{Analysis and Discussion}
From the results of Experiment~I and II (cf. Figure~\ref{fig:exp1_and_exp2_performance_user_aggregate}) we can observe that our RRF model ensembles achieve an f1 score improvement of roughly 85\% and 57\% over the baseline using Approach~1 and 2, respectively. RRF also compares favorably against the state-of-the-art classifiers in both experiments. 

Approach~1 is simpler to implement since it works at the Data Feed transaction level. Thus the predictions are computed in real-time directly for the user transactions observed in the Data Feed. The overall user experience, at any given point in time, can be computed by aggregating the predictions collected until that moment on a per user basis, e.g., by majority voting as we did in Experiment~I. In addition, the trees generated by this approach provide useful information on the relevance of the individual variables for classification (cf. Table~\ref{tab:exp1:variable_importance}), which can be used to support telco's insights on the most important dimensions, e.g., applications, mobile devices, or network KPIs, that impact the user experience.

For real-time prediction under Approach~II we first need to aggregate and summarize transactions per user, and then keep an up-to-date version of this \emph{user profile}. Then, these profiles can be used to assess the user experience at any time. Models learned using Approach~II can easily associate problems related to specific applications. For instance, the \emph{Apple Maps retransmission rate}, which corresponds to a node in the decision tree shown in Figure~\ref{fig:trees_experiment2}, identifies the cause of a major issue experienced in the telco network, which triggered a large number of customer care calls, both during the monitored days, and for two subsequent weeks, as confirmed by the operator. This is a promising result indicating that if the telco had deployed the proposed solution in a real-time manner, it would have been able to identify such an issue before an avalanche of calls are placed to the care center.

In future work we plan to explore a hybrid strategy that unifies our two approaches, in which predictions for user experience will be made simultaneously and combined in real time. This strategy would leverage the advantages of each approach at capturing the user experience for short (transaction level) and long terms (user profile) simultaneously, as well as preserving the interpretability of the models. 

The low precision of our models reflects how challenging this task is. We are currently experimenting with other tree ensemble strategies, similar to random forests~\cite{elemStatLearn}, in order to improve precision, e.g., by relaxing the interpretability constraint. One possible direction would be to include in a model ensemble a one-class classifier to better address the unbalance observed in our dataset, i.e., only a limited number of users actually call the care center to report an issue. Observe that the improvements for f1 are largely thanks to recall, which also indicates an increase in the number of false positives. That is, the model falsely predicting that the user will call the care center. Within the proactive customer care use case the telco is still interested in exploring a sample of such false positives, since they might indicate reasons for poor experience even if the customer did not call the care center. We plan to explore a relevance ranking mechanism for such instances, e.g., based on user behavior, and present the telco agents only a short list of interesting instances to examine.

We also want to explore additional features to train our models and improve their performance. Temporal and seasonal features such as the day of the week, month, and time of the day are important and should be considered when analyzing data streams for longer periods than the one studied here, i.e., 5 days of data. Furthermore, modeling the quality and user perception of an app based on additional sources, such as the user rating given in app market places or social media opinions, is an interesting direction to consider.

The customer care logs available to us for this study are noisy and the reason associated with each call was either uncategorized or not precise enough to be modeled explicitly. This is a clear limitation of our study. However, it would be straightforward for telcos, that have full access to these reports and their history, to leverage this important information as part our approach.
%

Finally, we plan to explore in more detail the individual aspects of user behavior to build personalized models for users or segments of users directly from the data stream (e.g., as in ~\cite{diaz_aviles_stream_topn}), an approach that we expect will be able to predict more accurately the mobile experience of each subscriber.

\section{Conclusion and Future Work}
\label{sec:conclusion}
In this paper we have presented a novel approach to estimate in near-real-time the user experience for telco customers from a rich set of mobile phone activity data and a log of customer calls to the telco's care center. 

We presented the real-world solution to tackle this problem and carried out an empirical evaluation that shows that our approach achieves promising results that overall show the potential for measuring user experience at any point in time and without intrusively eliciting explicit feedback from customers.

Our approach for user experience prediction can be leveraged to improve churn forecasting models as well as to indirectly estimate network promoter scores (NPS), a metric widely used in the telco industry to estimate future revenue. 
\bibliographystyle{IEEEtran}
\bibliography{IEEEabrv,2015_ieee_bd} 

\begin{thebibliography}{10}
\providecommand{\url}[1]{#1}
\csname url@samestyle\endcsname
\providecommand{\newblock}{\relax}
\providecommand{\bibinfo}[2]{#2}
\providecommand{\BIBentrySTDinterwordspacing}{\spaceskip=0pt\relax}
\providecommand{\BIBentryALTinterwordstretchfactor}{4}
\providecommand{\BIBentryALTinterwordspacing}{\spaceskip=\fontdimen2\font plus
\BIBentryALTinterwordstretchfactor\fontdimen3\font minus
  \fontdimen4\font\relax}
\providecommand{\BIBforeignlanguage}[2]{{%
\expandafter\ifx\csname l@#1\endcsname\relax
\typeout{** WARNING: IEEEtran.bst: No hyphenation pattern has been}%
\typeout{** loaded for the language `#1'. Using the pattern for}%
\typeout{** the default language instead.}%
\else
\language=\csname l@#1\endcsname
\fi
#2}}
\providecommand{\BIBdecl}{\relax}
\BIBdecl

\bibitem{itu2014}
``{The World in 2014: ICT Facts and Figures},'' International Telecommunication
  Union (ITU), Tech. Rep., 2014.

\bibitem{2013:forbes}
A.~Swinscoe, ``{A Story About The Benefits Of Proactive Customer Service.
  Forbes.}'' \url{http://www.forbes.com/sites/adrianswinscoe/}, (accessed
  online: 2015-03).

\bibitem{cisco_vni_stats}
{Cisco}, ``{Visual Networking Index -- Mobile Forecast Highlights, 2014 –
  2019},''
  \url{http://www.cisco.com/c/dam/assets/sol/sp/vni/forecast_highlights_mobile},
  (accessed online: 2015-07).

\bibitem{ngai2009application}
E.~W. Ngai, L.~Xiu, and D.~C. Chau, ``Application of data mining techniques in
  customer relationship management: A literature review and classification,''
  \emph{Expert systems with applications}, 2009.

\bibitem{hung2006applying}
S.-Y. Hung, D.~C. Yen, and H.-Y. Wang, ``Applying data mining to telecom churn
  management,'' \emph{Expert Systems with Applications}, 2006.

\bibitem{wei2002turning}
C.-P. Wei and I.-T. Chiu, ``Turning telecommunications call details to churn
  prediction: a data mining approach,'' \emph{Expert systems with
  applications}, 2002.

\bibitem{richter2010predicting}
Y.~Richter, E.~Yom-Tov, and N.~Slonim, ``Predicting customer churn in mobile
  networks through analysis of social groups.'' in \emph{SDM}, 2010.

\bibitem{rowe2013mining}
M.~Rowe, ``Mining user lifecycles from online community platforms and their
  application to churn prediction,'' in \emph{ICDM}, 2013.

\bibitem{mozer2000predicting}
M.~C. Mozer, R.~Wolniewicz, D.~B. Grimes, E.~Johnson, and H.~Kaushansky,
  ``Predicting subscriber dissatisfaction and improving retention in the
  wireless telecommunications industry,'' \emph{IEEE Transactions on Neural
  Networks}, 2000.

\bibitem{zhao2005customer}
Y.~Zhao, B.~Li, X.~Li, W.~Liu, and S.~Ren, ``Customer churn prediction using
  improved one-class support vector machine,'' in \emph{Advanced data mining
  and applications}.\hskip 1em plus 0.5em minus 0.4em\relax Springer, 2005.

\bibitem{4287441}
J.~Reades, F.~Calabrese, A.~Sevtsuk, and C.~Ratti, ``Cellular census:
  Explorations in urban data collection,'' \emph{IEEE Pervasive Computing},
  2007.

\bibitem{Bogomolov:2014}
A.~Bogomolov, B.~Lepri, J.~Staiano, N.~Oliver, F.~Pianesi, and A.~Pentland,
  ``{Once Upon a Crime: Towards Crime Prediction from Demographics and Mobile
  Data},'' in \emph{ICMI}, 2014.

\bibitem{tan2000textual}
P.-N. Tan, H.~Blau, S.~Harp, and R.~Goldman, ``Textual data mining of service
  center call records,'' in \emph{KDD}, 2000.

\bibitem{ibm_nf}
{IBM}, ``{Now Factory},'' \url{https://ibm.biz/IBM_NowFactory}, (accessed
  online: 2015-07).

\bibitem{elemStatLearn}
T.~J. Hastie, R.~J. Tibshirani, and J.~H. Friedman, \emph{The elements of
  statistical learning: data mining, inference, and prediction}, ser. Springer
  series in statistics, 2009.

\bibitem{marz2015big}
N.~Marz and J.~Warren, \emph{Big Data: Principles and best practices of
  scalable realtime data systems}.\hskip 1em plus 0.5em minus 0.4em\relax
  Manning Publications Co., 2015.

\bibitem{2015:ibm:bi}
IBM, ``{BigInsights for Apache Hadoop},''
  \url{https://ibm.biz/IBM_BigInsights}, (accessed online: 2015-07).

\bibitem{2015:ibm:streams}
{IBM}, ``{Streams},'' \url{https://ibm.biz/IBM_Streams}, (accessed: 2015-07).

\bibitem{ensembleLearning:polikar:2006}
R.~Polikar, ``Ensemble based systems in decision making,'' \emph{Circuits and
  Systems Magazine, IEEE}, vol.~6, no.~3, Third 2006.

\bibitem{ensembleLearning:2009}
Z.-H. Zhou, ``Ensemble learning,'' in \emph{Encyclopedia of Biometrics}, 2009.

\bibitem{2015:ibm:bigr}
{IBM}, ``{Big R},'' \url{https://ibm.biz/IBM_BigR}, (accessed online: 2015-07).

\bibitem{2014:rpart}
T.~Therneau, B.~Atkinson, and B.~Ripley, \emph{rpart: Recursive Partitioning
  and Regression Trees}, 2014, version 4.1-8.

\bibitem{diaz_aviles_stream_topn}
E.~Diaz-Aviles, L.~Drumond, L.~Schmidt-Thieme, and W.~Nejdl, ``Real-time top-n
  recommendation in social streams,'' in \emph{RecSys}, 2012.

\end{thebibliography}
\end{document}